\documentclass{article}
\usepackage{spconf}
\usepackage{amsmath}
\usepackage{graphicx}
\usepackage{subfig}

\graphicspath{{Images/}}
\DeclareGraphicsExtensions{.pdf,.jpg,.png}

\title{Detecting muscle activation using ultrasound speed of sound inversion with deep learning}
\name{Micha Feigin$^{\star}$ \qquad Manuel Zwecker$^{\dagger\ddagger}$ \qquad Daniel Freedman$^{\star\star}$ \qquad Brian W. Anthony$^{\star}$}

\address{$^{\star}$ Department of Mechanical Engineering, Massachusetts Institute of Technology, USA \\
    $^{\dagger}$Department of Neurological Rehabilitation, The Chaim Sheba Medical Center, Tel Hashomer, Israel \\
    $^{\ddagger}$Sackler Faculty of Medicine, Tel Aviv University, Israel \\
    $^{\star\star}$Google Research, Israel
    }
\begin{document}
%\ninept
%
\maketitle
\begin{abstract}
Functional muscle imaging is essential for diagnostics of a multitude of musculoskeletal afflictions such as degenerative muscle diseases, muscle injuries, muscle atrophy, and neurological related issues such as spasticity. However, there is currently no solution, imaging or otherwise, capable of providing a map of active muscles over a large field of view in dynamic scenarios.

In this work, we look at the feasibility of longitudinal sound speed measurements to the task of dynamic muscle imaging of contraction or activation. We perform the assessment using a deep learning network applied to pre-beamformed ultrasound channel data for sound speed inversion.

Preliminary results show that dynamic muscle contraction can be detected in the calf and that this contraction can be positively assigned to the operating muscles. Potential frame rates in the hundreds to thousands of frames per second are necessary to accomplish this.
\end{abstract}

% -----------------
\section{Introduction}
\label{sec:intro}
% -----------------

Muscle contraction and concomitant muscle force may be reduced in various musculoskeletal and neuromuscular disorders such as traumatic brain injury (TBI), cerebral palsy (CP) and multiple sclerosis (MS). For a long time, indirect quantification of dynamic muscle contraction has been performed on the joint level (range of motion and joint kinematics). Intrinsic parameters such as individual muscle force and moment, as well as joint load,  have been estimated from in vivo measurements using biomechanical analysis of musculoskeletal models. In recent years, there have been increasing research efforts in developing dynamic measures of musculo-tendon mechanical properties during movement using ultrasound (US) based methods, visible in real-time sonography. Elastography methods can be used to objectively assess muscle viscoelastic properties. Further, assessment of muscle contraction velocity enables us to precisely determine the muscle contraction characteristics (eccentric versus concentric) at different time points during movement so that an individual muscle's function as an actuator, decelerator, or a stabilizer for that particular movement can be better understood. Muscle contraction velocity may provide evidence about muscle heterogeneity, which has not been extensively examined in vivo. In particular, muscle deformation across different dynamic behaviors, reflecting uniform or non-uniform muscle characteristics may be useful to quantify disparities between healthy and pathological conditions. The architectural changes in muscle that have been studied during static and dynamic contractions include muscle thickness, fiber pennation, and fascicle length. The nature of the relationships between changes in force and muscle architecture varies between muscles and cannot be assumed. US has the potential to measure contractile ability if the nature of the relationship between changes in contraction level and muscle dimensions is elucidated.

We look at the problem through an elastic perspective. When muscles contract they stiffen. The Young's modulus is one of the physical property most closely related to our intuitive notion of stiffness.  Ultrasound shear wave elastography (SWE) is a method in medical ultrasound that aims to estimate this value. Consequently, the medical ultrasound research community has used SWE to assess muscle contraction and muscle force \cite{Shinohara2010,Hug2015,Brandenburg2016,Ryu2017,Lima2017a}. However, reliable results require imaging longitudinally to the muscles (i.e. with the probe oriented along the muscle fibers), thus not providing a cross-sectional image of muscle activation.

The bulk of this work has also been limited to isometric (static) muscle imaging. SWE generally suffers from extremely low frame rates and high sensitivity to both sonographer and subject movement. This makes the method unsuitable for assessing muscle activation during isotonic or other motion, even more so, with large fields of view of muscle cross-sections. Due to power limitations, SWE is also limited to high-end devices.

The closest result is the application of SWE to myocardial imaging \cite{Hsu2007}. These results however also present the physical limit of the technology, where due to limitations on tissue and probe heating, and the time to activate each source, frame rate reached for a 1.4 cm field is 10Hz.

Related work has looked into using ultrasound strain imaging \cite{Pigula2016,Witte2006}. This requires capturing multiple frames at the same point of view using different external loads on the ultrasound probe, making it unsuitable for dynamic imaging as well. Preliminary results using electro impedance tomography \cite{Silva2012,Murphy2018} do not extend to dynamic imaging either. 

We opt to look at longitudinal speed of sound (SoS). SoS depends on Young's modulus in a similar way to the shear speed of sound. It also depends on the bulk modulus, which is expected to increase during muscle contraction as well.

\begin{figure}
    \centering
    ~\hfill%
    \subfloat[b-mode]{\includegraphics[width=0.44\columnwidth]{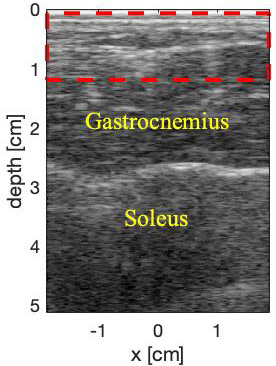}}%
    \hfill%
    \subfloat[Channel data]{\includegraphics[width=0.44\columnwidth]{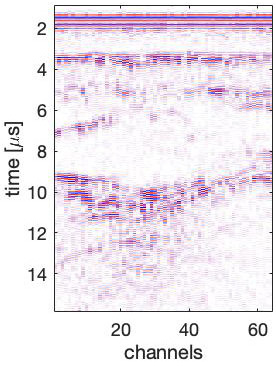}}
    \hfill~%

    \caption{B-Mode image of the calf muscles (a), the soleus and gastorcnemius. We opt to work with raw US channel data (b - cropped) instead as most cues required for SoS inversion are discarded as part of the image formation step.}
    \label{fig:input}
\end{figure}

In this work, we explore the feasibility of SoS inversion to the application of assessing muscle activation and functional muscle imaging based on modifications to a previously presented deep learning framework for sound speed inversion \cite{Feigin2019}. The SoS inversion network is applied to the pre-beamformed raw RF channel data generated from a single acoustic plane wave (Fig.~\ref{fig:input}). Cues regarding the sound speed of the domain through which the US pulse has propagated are hidden in the measured signal. However, most of this information is discarded as part of the classic delay and sum and envelope detection steps of the b-mode imaging process.

Deep learning has seen increasing popularity and success in the fields of vision and medical imaging for such tasks as classification, detection, and segmentation. However, only recently, deep learning has started being applied to physics and inverse problems \cite{Raissi2017b} and the imaging pipeline \cite{Perdios2019, Nair2018}. A neural network can be viewed as a black box performing non-linear regression with a set of parameters that control the relationship of the output on the input. The structure of the network affects both training properties as well as regularization, through the expressibility of the network. In this case, we wish to train the network to take raw channel data as input and produce sound speed maps as output. This is done by training it using simulation data. 

% -----------------
\section{Methods}
\label{sec:method}
% -----------------

\begin{figure}
    \centering
    \includegraphics[width=0.76\columnwidth]{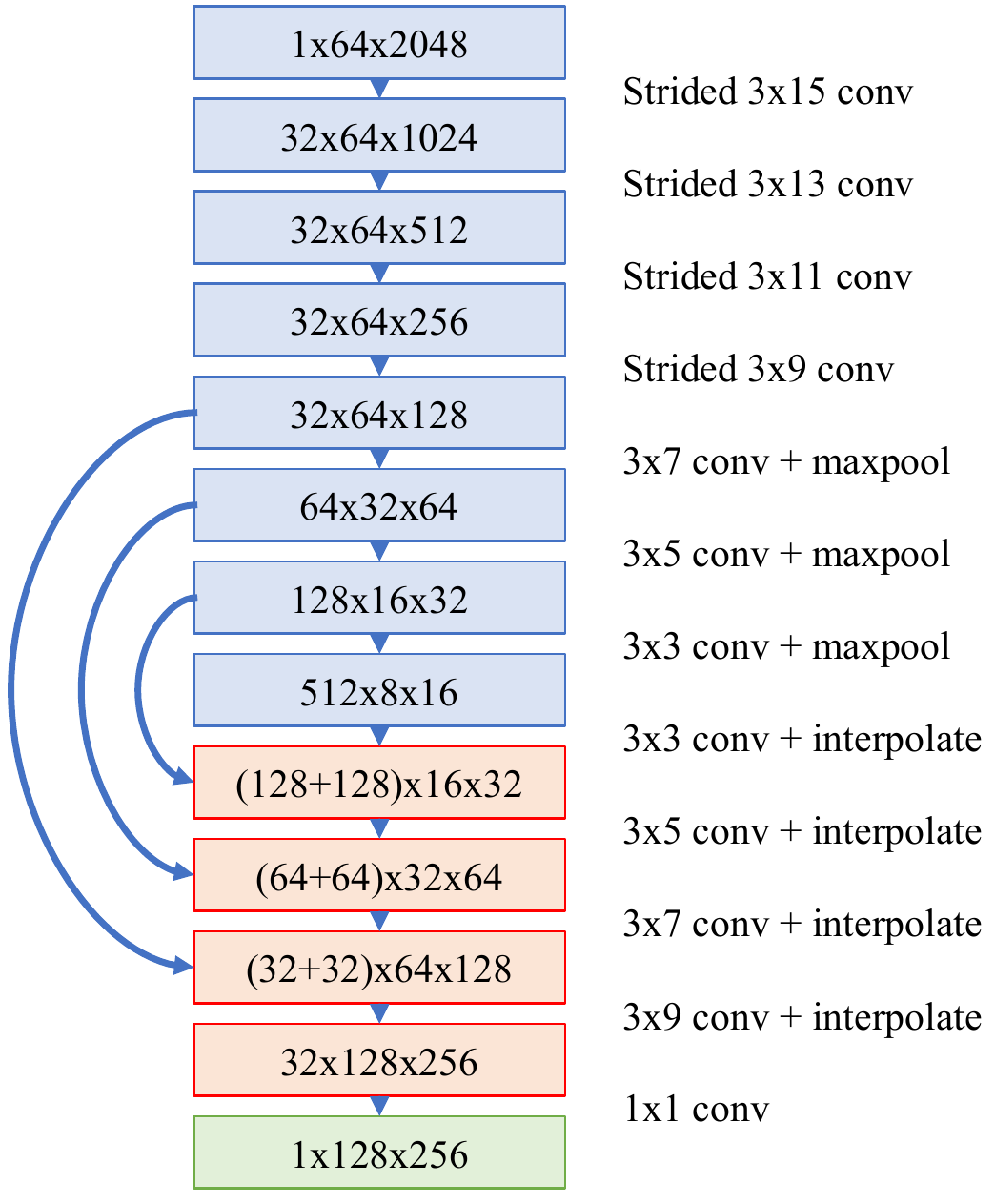}
    \caption{Topology of the deep neural network used in this word. Blue denotes the down sampling steps, orange the up sampling steps and green the output stage. Arrows show skip connections used.}
    \label{fig:network}
\end{figure}

For this work, we modified the network presented in \cite{Feigin2019}. Our modifications are presented in Fig.~\ref{fig:network}. For sound speed inversion, we use a fully convolutional neural network in an encoder-decoder topology, taking as input the raw RF ultrasound channel data, 64 channels by 2048 samples resulting from a single plane wave transmit. The network compresses the signal down to 8 by 16 samples with 512 features and expands it back to produce a 128 by 256 sound speed map. Due to the 2:1 decimation factor at each step, a filter size that complies with Nyquist sampling requirements is used. At the top level, the filter size is 3 channels by 15 samples, gradually decreasing in size to 3 channels by 3 samples, and increasing again in the upsampling steps to 3 by 9. This improved previous results while also removing block artifacts presenting with smaller filter sizes. Skip connections were added to the three innermost layers.

Due to a lack of an existing expert system capable of generating ground truth speed of sound maps, it is practically impossible to generate a large (or in fact, any) classified dataset on real data for training purposes. The network was thus trained on simulation data. We generated 6000 training samples and 800 testing samples by randomly generating between 1 and 5 spheres over the domain. Spheres were used due to simplicity, while still being close enough in shape to most organs. Attenuation and density were kept constant at 2.5 dB/cm and 0.9 gram/cm$^3$. Sound speed varied between 1300  m/s and 1800 m/s. Random 
speckle noise was added to the data as well as quantization and random Gaussian noise.

% ----------------------------
\section{Experimental results}
\label{sec:results}
% ----------------------------

\begin{figure}
    \centering
    \hfill%
    \subfloat[Anatomy]{\includegraphics[width=0.52\columnwidth]{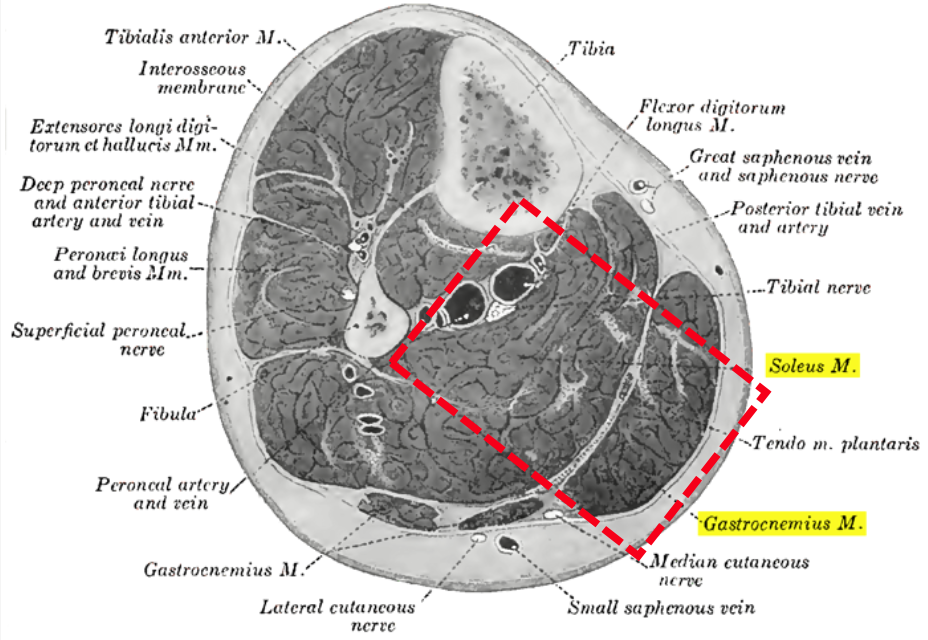}}%
    \hfill%
    \subfloat[Squat]{\includegraphics[width=0.32\columnwidth]{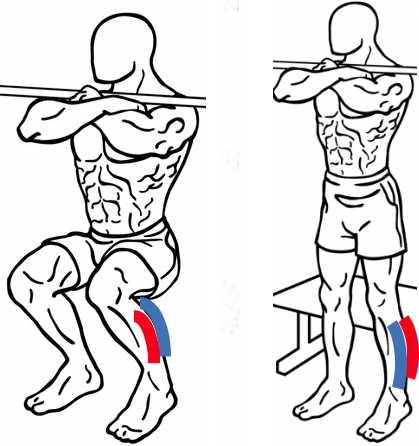}}%
    \hfill%
    
    \caption{Experimental setup. Image (a) shows the cross-section of the imaging anatomy, the calf muscles. We image the gastrocnemius and soleus muscles from a posterior view. The subject performs a squat maneuver, transitioning from standing to kneeling and back while standing on tip-toes. We expect the outer gastrocnemius to be active (red) in the standing position and the soleus to be relaxed (blue), and vice versa in the kneeling position.}
    \label{fig:experiment}
\end{figure}

For this work, we wish to show that we can differentiate between active and non-active or relaxed muscles. To this end we chose to look at the soleus and gastrocnemius (GC) muscles locates on the backside of the calf from a posterior view, as shown in Fig.~\ref{fig:experiment}. This was done for two reasons (1) we can see both muscles next to each other in the same ultrasound frame (2) the two muscles are active at different joint positions. This allows us to see the activation pattern shift from one muscle to the other. 

The two muscles are responsible for performing plantar flexion of the foot (with the GC also flexing the knee). Both muscles are connected at the Achilles tendon on the lower end. On the upper end, the GC connects to the femur while the soleus connects just below the knee. As a result, the GC performs most of the work with a straight leg while the soleus activates with a bent knee, as the GC deactivates due to the shortening of the muscle.

\begin{figure*}
    \centering
    \subfloat[Down phase]{\includegraphics[width=1\linewidth]{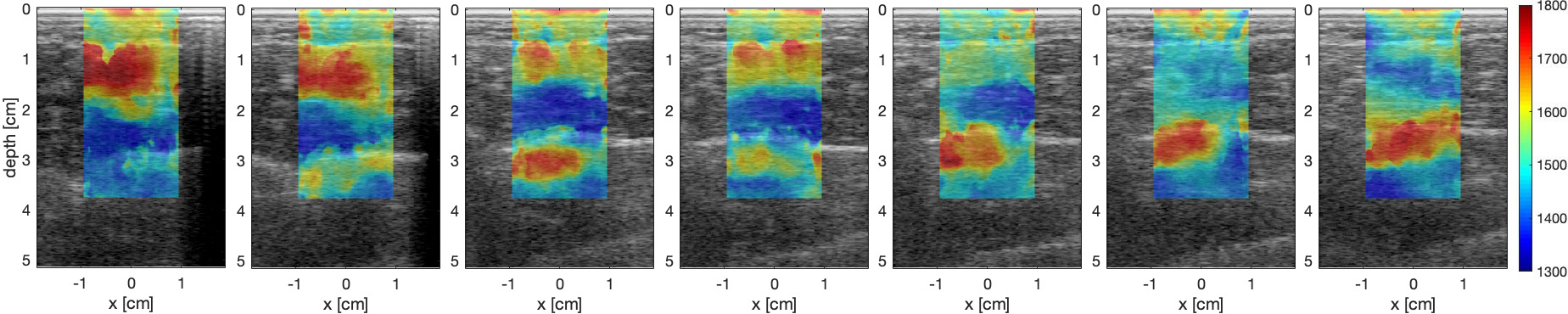}}

    \subfloat[Up phase]{\includegraphics[width=1\linewidth]{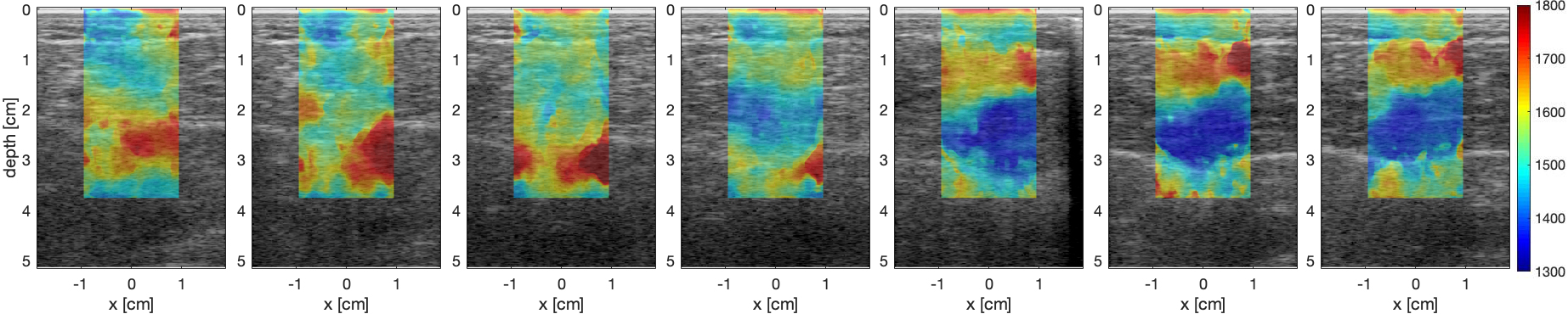}}
    
    \caption{Results for a squat maneuver. (a) shows the results for the transition from standing (left) to crouching (right) and (b) for the transition from crouching to standing position. Fig.~\ref{fig:input}(a) shows the location of the muscles in the frame.  Blue denotes lower sound speeds, correlated with relaxed muscles, red denotes higher sound speeds, correlated with contracted muscles. Both cases show that the soleus (inside of the leg - bottom of the image) is doing most of the work in the crouching position while the gastrocnemius (outside of the leg - top of the image) is doing most of the work in the standing position.}
    \label{fig:results}
\end{figure*}

To image the work transitioning between the two muscles, we asked a healthy human subject to perform a squat maneuver, i.e. transition from a standing position to a crouching position, and back to a standing position, while standing on tiptoes (Fig.~\ref{fig:experiment}). Experiments were carried using a protocol approved by the MIT Committee on the Use of Humans as Experimental Subjects (COUHES). Data was collected with a 128 element linear probe with a center frequency of 5 MHz using a Cephasonics Cicada ultrasound system. 

Our experimental results are presented in Fig.~\ref{fig:experiment}, with Fig.~\ref{fig:experiment}(a) showing the response for the transition from standing to crouching (down phase) and Fig.~\ref{fig:experiment}(b) for the transition from crouching to standing. Red color denotes higher sound speeds that correlate with contracted muscle, while blue color denotes lower sound speeds, that correlate with relaxed muscle.

Results in both cases match the expected action. In the down phase, we see the task start with the GC (outside of the leg / top part of the image) with the soleus starting to activate on the third frame, and slowly transition to the soleus, where we see that the GS slowly relaxes and the soleus slowly contracts. On the up phase, we see the complementary behavior, where the work is initially carried out by the soleus, and slowly transitions to the GC.

While the sound speeds we are currently recovering are out of the expected range, due to limitations on domain transfer of 2D simulations based training to real work measurements, the functional results show that the method is extremely capable at functional muscle imaging with strong potential for improving the results farther.

% ----------------------------
\section{Conclusion}
\label{sec:conclusions}
% ----------------------------

In this work, we have presented a longitudinal speed of sound based technique for dynamic functional muscle imaging. Results are based on a simulation-based deep learning method for sound speed inversion in ultrasound as applied to plane wave channel data.

Highly encouraging results show differentiation of muscle activation with potential frame rates of hundreds to thousands of frames per second.

This novel technique has the potential to assess dynamical muscle contraction with high resolution and capability of muscle delineation. 

\bibliographystyle{IEEEbib}
\bibliography{bibliography}

\end{document}